\documentclass[10pt]{iopart}
\usepackage{graphicx}

\usepackage[american]{babel}
\usepackage{amssymb}
\usepackage{amsfonts}
\begin{document}
\newtheorem{corollary}{Corollary}[section]
\newtheorem{remark}{Remark}[section]
\newtheorem{definition}{Definition}[section]
\newtheorem{theorem}{Theorem}[section]
\newtheorem{proposition}{Proposition}[section]
\newtheorem{lemma}{Lemma}[section]
\newtheorem{help1}{Example}[section]
\renewcommand{\theequation}{\arabic{section}.\arabic{equation}}

\title{Lower and upper estimates on the excitation threshold for breathers in DNLS lattices}
\author{J. Cuevas $^1$, N.I. Karachalios $^2$ and F. Palmero $^3$}

\address{
$^1$ Departamento de F\'{\i}sica Aplicada I, Escuela Universitaria
    Polit\'{e}nica,\\
    C/ Virgen de Africa, 7, University of Sevilla,\\
    41011 Sevilla, Spain\\
$^2$ Department of Mathematics, University of the Aegean,\\
  Karlovassi, 83200 Samos, Greece\\
$^3$ Departamento de F\'{\i}sica Aplicada I, ETSI Inform\'atica,\\
  Avd. Reina Mercedes s/n, University of Sevilla,\\
  41012 Sevilla, Spain \\
}

\eads{\mailto{jcuevas@us.es}, \mailto{karan@aegean.gr} and \mailto{palmero@us.es}}

\begin{abstract}
We propose analytical lower and upper estimates on the excitation threshold for breathers (in the form of spatially localized and time periodic solutions) in DNLS lattices with power nonlinearity. The estimation depending explicitly on the lattice parameters, is derived by a combination of a comparison argument on appropriate lower bounds depending on the frequency of each solution with a simple and justified heuristic argument. The numerical studies verify that the analytical estimates can be of particular usefulness, as a simple analytical detection of the activation energy for breathers in DNLS lattices.
\end{abstract}

\submitto{\JPA}
\maketitle

\section{Introduction}
A great deal of attention has been paid to the study of localization phenomena in nonlinear discrete systems in recent years, interest which has been summarized in a number of recent reviews \cite{reviews}. This growth has been motivated not only by its intrinsic theoretical interest, but also by numerous applications in areas as the nonlinear optics of waveguide arrays \cite{optics}, Bose-Einstein condensates \cite{bec_reviews}, micro-mechanical models of cantilever arrays \cite{sievers}, or some models of the complex dynamics of the DNA  \cite{peyrard}.

In this framework, perhaps one of the the most prototypical model is the so-called, discrete nonlinear Schr{\"o}dinger equation (DNLS) \cite{dnls,Eil03}. DNLS may arise as a direct model, as a tight binding approximation, or even as an envelope wave expansion and, it could be possible to say that the DNLS is one of the most ubiquitous models in the nonlinear physics of dispersive, discrete systems \cite{Panos_book}.

Our aim in the present paper is to determine analytical lower and upper useful bounds for the formation of spatially localized and time periodic modes  in focusing DNLS lattices, called discrete breathers or also DNLS solitons, with power nonlinearity \cite{Pal08,Jesus09}.

Flach, Kladko and MacKay \cite{{FlachMac}}, addressed the existence of {energy thresholds} for the formation of discrete breathers in in one-, two and three-dimensional lattices. They defined  {\em as energy thresholds, the positive lower energy bounds possessed by discrete breather families (DB).} Their numerical findings and heuristic arguments considered a generic class of Hamiltonian systems  and showed that the energy
of a DB family has a positive lower bound for lattice dimension
$N$ greater than or equal to some critical dimension $N_c$, whereas for
$N<N_c$ the energy goes to zero as the amplitude goes to
zero.

For the focusing DNLS
equation in the infinite lattice $\mathbb{Z}^N$,
\begin{eqnarray}
\label{focDNLS}
\mathrm{i}\dot{\psi}_n+ \epsilon(\Delta_d\psi)_n + \Lambda_n|\psi_n|^{2\sigma}=0,\;\;\Lambda>0,\;\;\sigma>0,
\end{eqnarray}
where $n=(n_1,n_2,\ldots,n_N)\in\mathbb{Z}^N$, the hypothesis suggested by Flach  Kladko and MacKay was resolved by Weinstein \cite{Wein99}.
In (\ref{focDNLS}), $(\Delta_d\psi)_n$ stands for the $N$-dimensional discrete Laplacian
\begin{eqnarray}
\label{DiscLap}
(\Delta_d\psi)_{n\in\mathbb{Z}^N}=\sum_{m\in \mathcal{N}_n}\psi_m-2N\psi_n.
\end{eqnarray}
Here $\mathcal{N}_n$ denotes the set of $2N$ nearest neighbors of
the point in $\mathbb{Z}^N$ with label $n$. The parameter
$\epsilon>0$ is a discretization parameter
$\epsilon\sim h^{-2}$ with $h$ being the lattice spacing and $\Lambda>0$ is the parameter of anharmonicity.

The hypothesis of \cite{FlachMac} was resolved in \cite{Wein99} for breathers in the ansatz of time-periodic solutions
\begin{equation}
\label{TP1}
\psi_n(t)=e^{\mathrm{i}\Omega t}\phi_n,\;\;\Omega>0,
\end{equation}
spatially localized in the sense $$|\psi_n|\rightarrow 0,\;\;
\mbox{as}\;\;|n|\rightarrow\infty,$$
(here  $|n|=\max_{1\leq i\leq N}|n_i|$ for
$n=(n_1,n_2,\ldots,n_N)\in\mathbb{Z}^N$).

Solutions (\ref{TP1}) of (\ref{focDNLS}) satisfy the
infinite system of algebraic equations
\begin{eqnarray}
\label{TP2}
-\epsilon(\Delta_d\phi)_n +\Omega\phi_n-\Lambda|\phi_n|^{2\sigma}\phi_n
& =0,\;\; n\in\mathbb{Z}^N.
\end{eqnarray}
We can associate a {\em power} to any solution of the form
(\ref{TP1}), defined as
\begin{equation}
  \label{power}
  \mathcal{R}[\phi] = \sum_{n\in\mathbb{Z}^N}|\phi_n|^{2}.
\end{equation}
The power (\ref{power}) together with the Hamiltonian
\begin{eqnarray}
\label{WeinA2}
\mathcal{H}[\phi]&=\epsilon(-\Delta_d\phi,\phi)_2-\frac{1}{\sigma+1}
\sum_{n\in\mathbb{Z}^N}|\phi_n|^{2\sigma+2},
\end{eqnarray}
are the fundamental conserved quantities for (\ref{focDNLS}).

For the proof of \cite{Wein99} on the existence of the excitation threshold, a discrete version of a Sobolev-Gagliardo-Nirenberg inequality is crucial. This discrete version reads as
\begin{eqnarray}
\label{WGN1}
\sum_{n\in\mathbb{Z}^N}|\phi_n|^{2\sigma+2}\leq
C\left(\sum_{n\in\mathbb{Z}^N}| \phi_n|^2\right)^{\sigma}(-\Delta_d\phi,\phi)_2,\;\;\sigma\geq\frac{2}{N},
\end{eqnarray}
If $C_*$ is the infimum over all such constants for which inequality
(\ref{WGN1}) holds, then the excitation threshold
$\mathcal{R}_{\mathrm{thresh}}$ is defined by \cite[pg. 680, Eqn.
(4.2)]{Wein99}
\begin{eqnarray}
\label{WGN2}
(\sigma+1)\epsilon\left(\mathcal{R}_{\mathrm{thresh}}\right)^{-\sigma}=C_*,
\end{eqnarray}
and the optimal constant $C_*$ has the variational characterization
\begin{eqnarray*}
\frac{1}{C_*}=
\inf_{
\begin{array}{c}
\phi \in \ell^2\\
\phi \neq 0
\end{array}}\frac{\left(\sum_{n\in\mathbb{Z}^N}|\phi_n|^2\right)^{\sigma}
(-\Delta_d\phi,\phi)_2}{\sum_{n\in\mathbb{Z}^N}|\phi_n|^{2\sigma+2}}.
\end{eqnarray*}
Then Weistein's result on the excitation threshold reads as follows: if
$\mathcal{R}>\mathcal{R}_{\mathrm{thresh}}$ then $\mathcal{I}_{\mathcal{R}}<0$, and a ground state breather exists, that is,  minimizer of the variational problem
\begin{eqnarray}
\label{WeinA1}
\mathcal{I}_{\mathcal{R}}=\inf\left\{\mathcal{H}[\phi]\;:\;
  \mathcal{P}[\phi] =\mathcal{R}\right\}.
\end{eqnarray}
On the other hand, if if $\mathcal{R}<\mathcal{R}_{\mathrm{thresh}}$ then
$\mathcal{I}_{\mathcal{R}}=0$, and there is no ground state minimizer of (\ref{WeinA1}). In the light of the results of Weinstein, the critical dimension predicted by Flach, Kladko and MacKay is defined for the DNLS (\ref{focDNLS}) as
\begin{eqnarray}
\label{critDim}
N_c=\frac{2}{\sigma}.
\end{eqnarray}

In this paper we propose analytical lower and upper estimates on the excitation threshold $\mathcal{R}_{\mathrm{thresh}}$, which are depending explicitly on the lattice parameters. The derivation of these  estimates in section 2, can be briefly described as follows: for the lower estimate, we derive first by a fixed point argument, a lower bound on the power of the  breather solution satisfied for any $\Omega>0$. The role of such local bounds (through their dependence on the frequency $\Omega$) as thresholds on the existence of breather solutions has been analyzed in detail and tested numerically in \cite{JCN,JCN2}. Then this is compared with a second local lower bound involving this time the unknown value of $\mathcal{R}_{\mathrm{thresh}}$. Although the lower bound for $\mathcal{R}_{\mathrm{thresh}}$ derived as above,  depends on an unspecified positive integer, its appropriate value can be easily determined by  a simple and justified heuristic argument, explained in detail in subsection 2.1. The derivation of the upper bound comes out by simply examining the interpolation inequality (\ref{WGN1}) in comparison with the standard embedding inequality between the $\ell^p$-sequence spaces.

The numerical studies performed in subsection 2.1, justify that the estimates for $\mathcal{R}_{\mathrm{thresh}}$ can be useful (on the account of their explicit  dependence on the lattice parameters and the simplicity of the formulas), in ``trapping'' the exact value of $\mathcal{R}_{\mathrm{thresh}}$ for the cases of nonlinearity exponent $\sigma$ and dimension $N$ which are of primary physical interest. This ``trapping'' is of particular interest in applications since the analytical estimation of the excitation threshold can be used  for a simple calculation of the activation energy needed for the experimental detection of discrete breathers \cite{FlachMac}. It is important to recall that the excitation threshold appears in the formal continuum limit $\epsilon\rightarrow\infty$ only in the case $\sigma=2/N$ \cite{Wein99}.

\section{Analytical lower and upper bounds for $\mathcal{R}_{\mathrm{thresh}}$.}
Our arguments on the determination of simple analytical bounds for the excitation threshold $\mathcal{R}_{\mathrm{thresh}}$ will be based on some technical lemmas involving lower bounds for the power of solutions (\ref{TP1}) \emph{for all} $\Omega>0$.
\begin{lemma}
\label{lem1}
The power of a nontrivial breather solution (\ref{TP1}) of (\ref{focDNLS}), satisfies the lower bound
\begin{eqnarray}
 \label{eq7}
\mathcal{R}_{\mathrm{min},1}:=\mathcal{R}_{\mathrm{thresh}}\cdot\left[\frac{\Omega}{4\epsilon\Lambda N(\sigma+1)}
\right]^{\frac{1}{\sigma}}<\mathcal{R}[\phi]\;\;\mbox{for all}\;\;\Omega>0.
\end{eqnarray}
\end{lemma}
\textbf{Proof:} Multiplying (\ref{TP2}) in the $\ell^2$-scalar product we infer that $\phi$ satisfies the energy equation
\begin{eqnarray}
\label{eq5}
{\epsilon}(-\Delta_d\phi,\phi)_{2}
+\Omega\sum_{n\in\mathbb{Z}^N}|\phi_n|^2
=\Lambda\sum_{n\in\mathbb{Z}^N}|\phi_n|^{2\sigma +2},\;\;\mbox{for all}\;\;\Omega>0.
\end{eqnarray}
Now inserting the inequality (\ref{WGN1}) in the right-hand side of (\ref{eq5}), and noting that
\begin{eqnarray}
 \label{eq5a}
(-\Delta_d\phi,\phi)_2\leq 4N\sum_{n\in\mathbb{Z}^N}| \phi_n|^2,
\end{eqnarray}
we deduce that
\begin{eqnarray*}
 {\epsilon}(-\Delta_d\phi,\phi)_{2}
+\Omega\sum_{n\in\mathbb{Z}^N}|\phi_n|^2
&\leq&\Lambda C_{*}\left(\sum_{n\in\mathbb{Z}^N}| \phi_n|^2\right)^{\sigma}(-\Delta_d\phi,\phi)_2\\
&\leq&4\epsilon\Lambda N C_{*}\left(\sum_{n\in\mathbb{Z}^N}| \phi_n|^2\right)^{\sigma+1}.
\end{eqnarray*}
Since $(-\Delta_d\phi,\phi)_2\geq 0$ we infer that
\begin{eqnarray}
 \label{eq6}
\Omega\sum_{n\in\mathbb{Z}^N}|\phi_n|^2\leq 4\Lambda NC_{*}\left(\sum_{n\in\mathbb{Z}^N}| \phi_n|^2\right)^{\sigma+1}.
\end{eqnarray}
By substitution of (\ref{WGN1}) into (\ref{eq6}), we derive the lower bound (\ref{eq7}).\ \ $\diamond$
\begin{lemma}
\label{lem2}
Let $\kappa\in\mathbb{R}^+$, $\kappa >\frac{1}{2}$, arbitrary. Then every  non-trivial breather solution (\ref{TP1}) of (\ref{focDNLS}) has power satisfying
\begin{eqnarray}
 \label{r1}
\mathcal{R}_{\mathrm{min},2}(\kappa):=\left[\frac{\sqrt{2\kappa-1}}{\kappa}\cdot\frac{\Omega}{\Lambda(2\sigma +1)}
\right]^{\frac{1}{\sigma}}<\mathcal{R}[\phi] \;\;\mbox{for all}\;\;\Omega>0.
\end{eqnarray}
\end{lemma}
\textbf{Proof:} We recall and modify appropriately  the fixed point argument of \cite{JCN, K1}. We consider the operator
\begin{eqnarray}
\label{r2}
-\epsilon\Delta_d+\Omega: \ell^2\rightarrow\ell^2.
\end{eqnarray}
being linear and continuous. It also satisfies  the assumptions  of Lax-Milgram Theorem \cite[Theorem 18.E, pg. 68]{zei85}: Note that
\begin{eqnarray}
\label{check}
\epsilon(-\Delta_d\phi,\phi)_2+\Omega ||\phi||^2_2\geq \Omega|\phi||^2_2\;\;\mbox{for all}\;\;\phi\in\ell^2.
\end{eqnarray}
Then according to Lax-Milgram theorem, for \emph{given} $z\in\ell^2$, the linear operator equation
\begin{eqnarray}
\label{linear}
-\epsilon\Delta_d\phi_n+\Omega\phi_n=\Lambda|z_n|^{2\sigma}z_n,\;\;\Lambda>0,
\end{eqnarray}
has a unique solution $\phi\in\ell^2$, since
\begin{eqnarray*}
|||z|^{2\sigma}z||^2_2\leq \sum_{n\in\mathbb{Z}^N}|z_n|^{4\sigma +2}
\leq ||z||_2^{4\sigma +2}.
\end{eqnarray*}
 Hence we are allowed to define the map
$\mathcal{P}:\ell^2\rightarrow\ell^2$, by $\mathcal{P}(z):=\phi$ where $\phi$ is the unique solution of the operator equation (\ref{linear}). Clearly the map $\mathcal{P}$ is well defined. Let $\zeta,\xi$ be in the closed ball
$$B_R:=\{z\in\ell^2\;:||z||_{\ell^2}\leq R\},$$
such that $\phi=\mathcal{P}(\zeta)$, $\psi=\mathcal{P}(\xi)$. The difference $\chi:=\phi-\psi$ satisfies the equation
\begin{eqnarray}
\label{claim2}
-\epsilon\Delta_d\chi_n+\Omega\chi_n=\Lambda(|\zeta_n|^{2\sigma}\zeta_n-|\xi_n|^{2\sigma}\xi_n)
\end{eqnarray}
We recall that for any $F\in \mathrm{C}(\mathbb{C},\mathbb{C})$ which
takes the form $F(z)=g(|\zeta|^2)\zeta$, with $g$ real and
sufficiently smooth, the following relation holds
\begin{eqnarray}
\label{GLTaylor12}
F(\zeta)-F(\xi)=\int_{0}^{1}\left\{(\zeta-\xi)(g(r)+rg'(r))
+(\overline{\zeta}-\overline{\xi})\Phi^2 g'(r)\right\}d\theta,
\end{eqnarray}
for any $\zeta,\;\xi\in \mathbb{C}$,where $\Phi=\theta \zeta
+(1-\theta)\xi$, $\theta\in (0,1)$ and $r=|\Phi|^2$ (see \cite[pg.
202]{GiVel96}).  Applying (\ref{GLTaylor12}) for the case of
$F(\zeta)=|\zeta|^{2\sigma}\zeta$, one finds that
\begin{eqnarray}
\label{Taylor13}
|\zeta|^{2\sigma}\zeta-|\xi|^{2\sigma}\xi=\int_0^1[(\sigma
+1)(\zeta-\xi)|\Phi|^{2\sigma}
+\sigma(\overline{\zeta}-\overline{\xi})\Phi^2|\Phi|^{2\sigma
-2}]d\theta.
\end{eqnarray}
Assuming that $\zeta$, $\xi\in \mathcal{B}_R$, and noting that
$||\Phi||_2\leq R$, we get from (\ref{Taylor13}) the inequality
\begin{eqnarray}
\label{Taylor14}
\sum_{n\in\mathbb{Z}^N}||\zeta_n|^{2\sigma}\zeta_n-|\xi_n|^{2\sigma}\xi_n|^2&\leq& (2\sigma+1)^2\sum_{n\in\mathbb{Z}^N}
\left\{\int_0^1|\Phi_n|^{2\sigma}|\zeta_n-\xi_n|d\theta\right\}^2\nonumber\\
&\leq&
(2\sigma+1)^2\sum_{n\in\mathbb{Z}^N}\left\{\int_0^1||\Phi||_2^{2\sigma}|
\zeta_n-\xi_n|d\theta\right\}^2\nonumber\\
&\leq&
(2\sigma+1)^2\sum_{n\in\mathbb{Z}^N}\left\{\int_0^1 R^{2\sigma}|\zeta_n-\xi_n|
d\theta\right\}^2\nonumber\\
&&=(2\sigma +1)^2R^{4\sigma}\sum_{n\in\mathbb{Z}^N}|\zeta_n-\xi_n|^2.
\end{eqnarray}
Taking now the scalar product of (\ref{claim2}) with $\chi$ in $\ell^2$ and using (\ref{Taylor14}), we have
\begin{eqnarray}
\label{cmap1a}
\epsilon(-\Delta_d\chi,\chi)_2+\Omega ||\chi||^2_2&\leq& \Lambda||\chi||_2||\,|\zeta|^{2\sigma}\zeta-|\xi|^{2\sigma}\xi||_2\nonumber\\
&\leq&\Lambda (2\sigma +1)^2R^{2\sigma}||\chi||_2||\zeta-\xi||_2.
\end{eqnarray}
Applying next Young's inequality
$$ab<\frac{\hat{\epsilon}}{p}a^p+\frac{1}{q\hat{\epsilon}^{q/p}}b^q,\;\;\mbox{for any}\;\;\hat{\epsilon}>0, \;\;1/p+1/q=1,$$
with $p=q=2$, $a=||\chi||_2,\,b=||\zeta-\xi||_2$ and
\begin{eqnarray*}
 \hat{\epsilon}=\frac{\omega}{\kappa},\;\;\kappa\in\mathbb{R}^+,\;\;\kappa> 1/2,
\end{eqnarray*}
in (\ref{cmap1a}), we get that
\begin{eqnarray}
\label{claim4}
\frac{(2\kappa-1)\Omega}{2\kappa}||\chi||_2^2
\leq \frac{\kappa}{2\Omega}\Lambda^2 (2\sigma+1)^2R^{4\sigma}||\zeta-\xi||^2_2.
\end{eqnarray}
From (\ref{claim4}), we conclude with
\begin{eqnarray*}
||\chi||_2^2=||\mathcal{P}(z)-\mathcal{P}(\xi)||^2_2
\leq \frac{\kappa^2}{\Omega^2(2k-1)}\Lambda^2 (2\sigma+1)^2R^{4\sigma}||\zeta-\xi||^2_2.
\end{eqnarray*}

Since $\mathcal{P}(0)=0$, from inequality  (\ref{claim4}) we derive that the map  $\mathcal{P}:B_R\rightarrow B_R$ is a Lipschitz map with Lipschitz constant
$$L=\frac{\kappa}{\Omega\sqrt{2k-1}}\Lambda (2\sigma+1)R^{2\sigma}.$$

The map $\mathcal{P}$ will be a contraction and will have a unique fixed point if $L<1$. This unique fixed point will be the trivial one, since $\mathcal{P}(0)=0$.  Hence, for $$R^2<\left[\frac{\sqrt{2\kappa-1}}{\kappa}\cdot\frac{\Omega}{\Lambda(2\sigma +1)}
\right]^{\frac{1}{\sigma}}$$
the only breather solution is the trivial. Therefore, a non-trivial breather solution (\ref{TP1}) should have power $\mathcal{R}[\phi]\geq \mathcal{P}_{\mathrm{min},2}$. \ \ $\diamond$

Keeping the positive constant $\kappa>1/2$ undetermined is crucial to derive a $\kappa$-dependent lower bound for $\mathcal{R}_{\mathrm{thresh}}$. However the numerical investigations, will reveal that for practical purposes the constant $\kappa$ can be easily determined. We start by proving the following
\begin{proposition}
\label{prop1}
Let $\sigma\geq 2/N$.  There exist $\kappa_{\mathrm{crit}}>1/2$ such that
 \begin{eqnarray}
\label{res1}
\left[\frac{\sqrt{2\kappa_{\mathrm{crit}}-1}}{\kappa_{\mathrm{crit}}}\cdot\frac{4N\epsilon(\sigma+1)}{2\sigma+1}
\right]^{\frac{1}{\sigma}}<R_{\mathrm{thresh}}<\left[4\epsilon N(\sigma+1)\right]^{\frac{1}{\sigma}}.
\end{eqnarray}
\end{proposition}
\textbf{Proof:} It follows from  Lemma \ref{lem1} that $\lim_{\kappa\rightarrow\infty}\mathcal{R}_{\mathrm{min},2}(\kappa)=0$, therefore, we can make  $\mathcal{R}_{\mathrm{min},2}(\kappa)$  as small as we please by taking $\kappa$ large enough. Thus, there exists a $\kappa_{\mathrm{crit}}$ such that
$\mathcal{R}_{\mathrm{min},2}(\kappa_{\mathrm{crit}})<\mathcal{R}_{\mathrm{min},1}$, implying the left-hand side of (\ref{res1}). For the right-hand side of (\ref{res1}), we recall first that
\begin{eqnarray}
 \label{com1}
\sum_{n\in\mathbb{Z}^N}| \phi_n|^p\leq \left(\sum_{n\in\mathbb{Z}^N}| \phi_n|^q\right)^{\frac{p}{q}},\;\;\mbox{for all}\;\; 1\leq q\leq p\leq\infty.
\end{eqnarray}
Applying the inequality (\ref{com1}) for $p=2\sigma+2$ and $q=2$ we get that
\begin{eqnarray*}
\sum_{n\in\mathbb{Z}^N}|\phi_n|^{2\sigma+2}\leq \left(\sum_{n\in\mathbb{Z}^N}| \phi_n|^2\right)^{\sigma+1},\;\;\mbox{\emph{for all}}\;\;\sigma\geq 0,\;\;\phi\in\ell^2.
\end{eqnarray*}
From (\ref{WGN1b}) we have
\begin{eqnarray*}
\sup_{
\begin{array}{c}
\phi \in \ell^2\\
\phi \neq 0
\end{array}}\frac{\sum_{n\in\mathbb{Z}^N}|\phi_n|^{2\sigma+2}}
{\left(\sum_{n\in\mathbb{Z}^N}|\phi_n|^2\right)^{\sigma+1}}\leq 1,\;\;\mbox{\emph{for all}}\;\;\sigma\geq 0.
\end{eqnarray*}
Since the above inequality holds for {\em all $\sigma\geq 0$ and all $\phi\in\ell^2$}, by setting $\phi=\mathbf{e}^j$ for arbitrary $j\in\mathbb{N}$, any element of the orthonormal basis of $\ell^2$, in (\ref{com1}),  we get that
\begin{eqnarray}
\label{WGN1b}
\sup_{
\begin{array}{c}
\phi \in \ell^2\\
\phi \neq 0
\end{array}}\frac{\left(\sum_{n\in\mathbb{Z}^N}|\phi_n|^2\right)^{\sigma+1}}
{\sum_{n\in\mathbb{Z}^N}|\phi_n|^{2\sigma+2}}=1,\;\;\mbox{\emph{for all}}\;\;\sigma\geq 0.
\end{eqnarray}
On the other hand, it follows from (\ref{WGN1}) and (\ref{eq5a}) that
\begin{eqnarray*}
\sum_{n\in\mathbb{Z}^N}|\phi_n|^{2\sigma+2}\leq
C_*\left(\sum_{n\in\mathbb{Z}^N}| \phi_n|^2\right)^{\sigma}(-\Delta_d\phi,\phi)_2&\leq& 4NC_*\left(\sum_{n\in\mathbb{Z}^N}| \phi_n|^2\right)^{\sigma}\sum_{n\in\mathbb{Z}^N}| \phi_n|^2.\nonumber\\
&=&4NC_*\left(\sum_{n\in\mathbb{Z}^N}| \phi_n|^2\right)^{\sigma+1},\;\;\sigma\geq 2/N,
\end{eqnarray*}
which implies that
\begin{eqnarray}
\label{WGN1a}
 \frac{\left(\sum_{n\in\mathbb{Z}^N}|\phi_n|^2\right)^{\sigma+1}}
{\sum_{n\in\mathbb{Z}^N}|\phi_n|^{2\sigma+2}}\leq 4NC^*,\;\;\mbox{for all}\;\;\sigma\geq 2/N ,\;\;\phi\in\ell^2.
\end{eqnarray}
Then a comparison of (\ref{WGN1b}) which holds {\em for all} $\sigma>0$, with (\ref{WGN1a}) implies that
\begin{eqnarray}
\label{implb}
1<4NC_*=4\epsilon N(\sigma+1)R_{\mathrm{thresh}}^{-\sigma},
\end{eqnarray}
from which we conclude the right-hand side of (\ref{res1}). Even in a simpler way, one can set in (\ref{WGN1a}) any element of the orthonormal basis of $\ell^2$ to derive (\ref{implb}). \ \ $\diamond$
\section{Numerical study}
\begin{figure}
\begin{center}
    \begin{tabular}{cc}
    \includegraphics[scale=0.45]{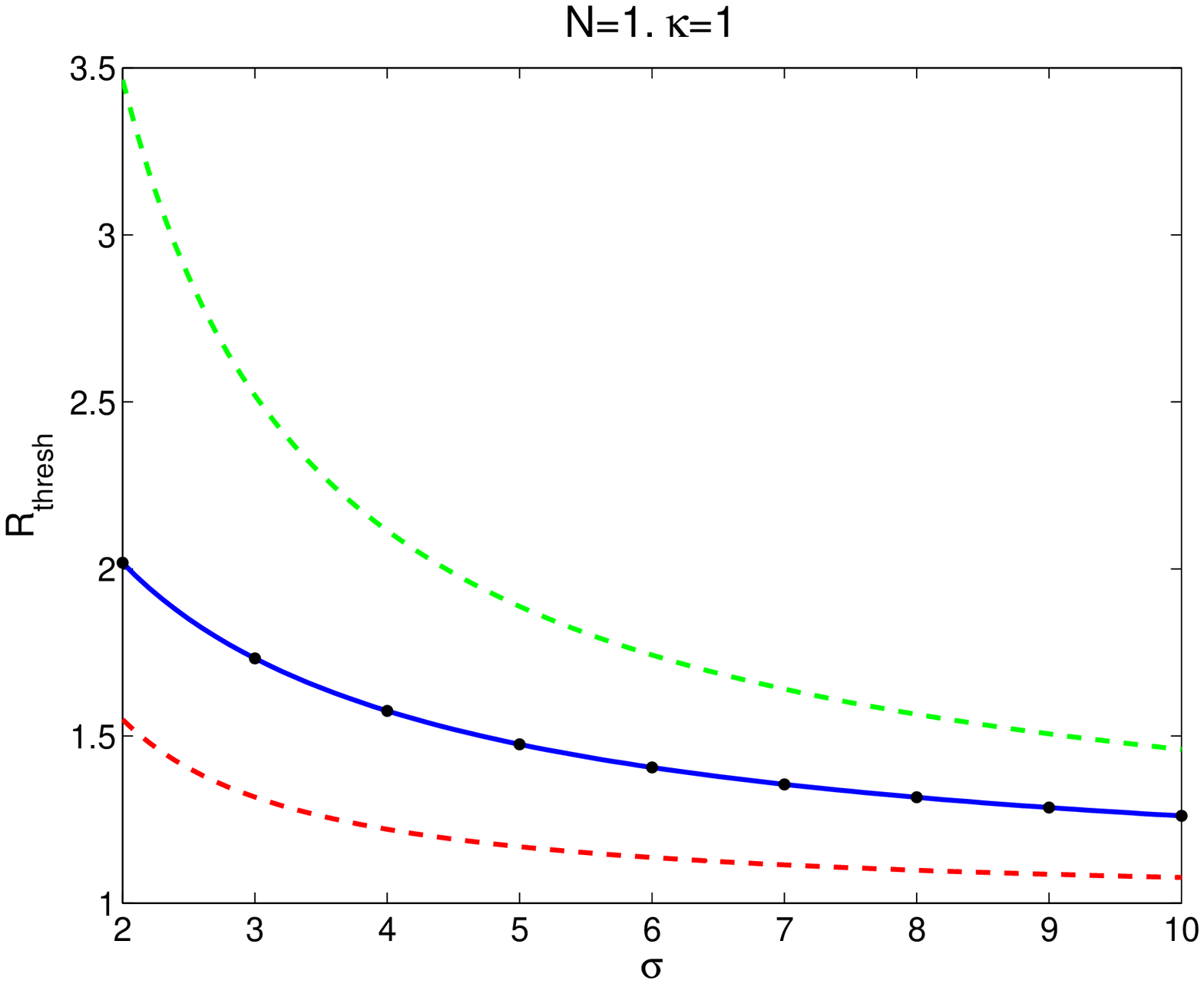} &
    \includegraphics[scale=0.45]{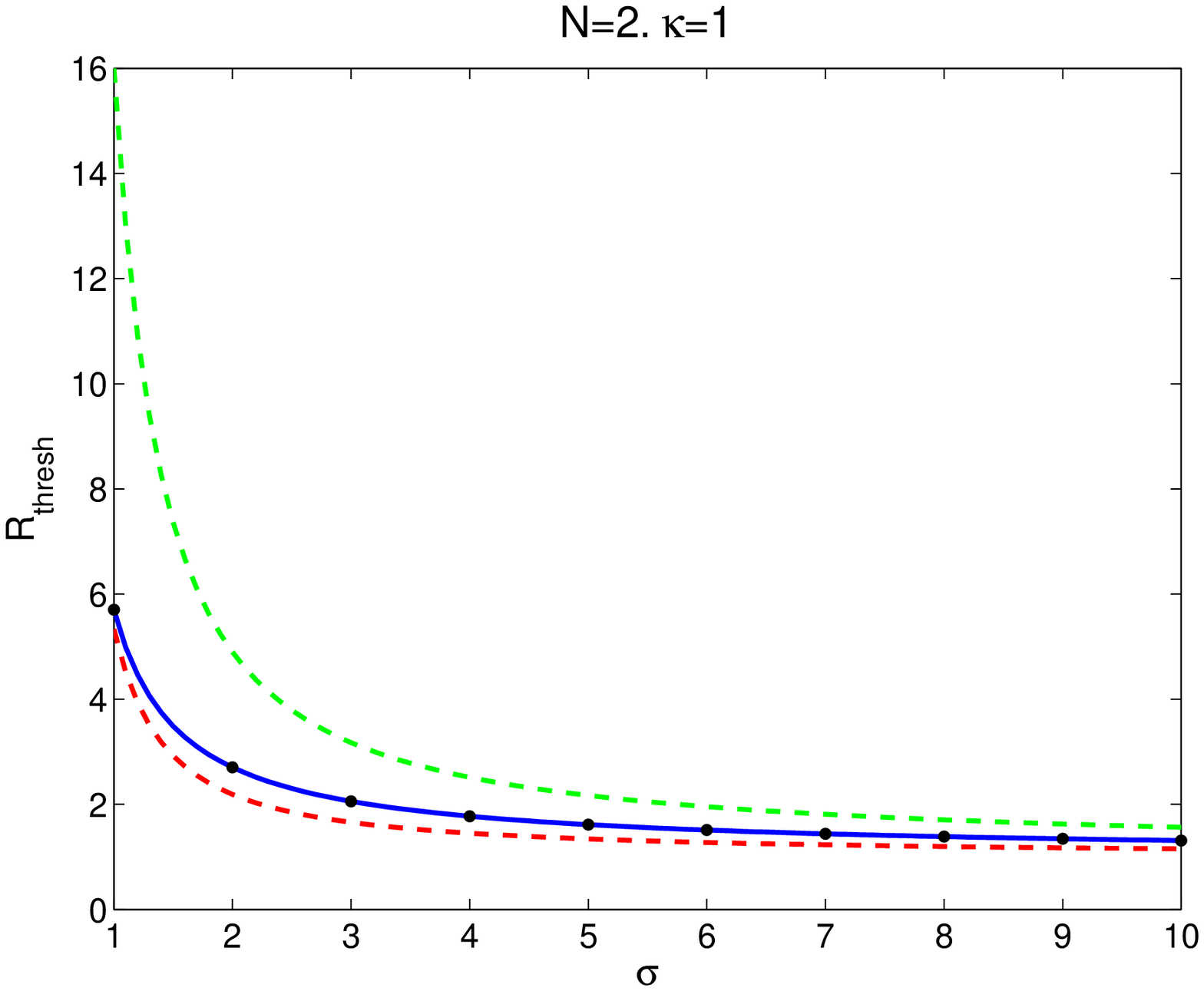} \\
    \includegraphics[scale=0.45]{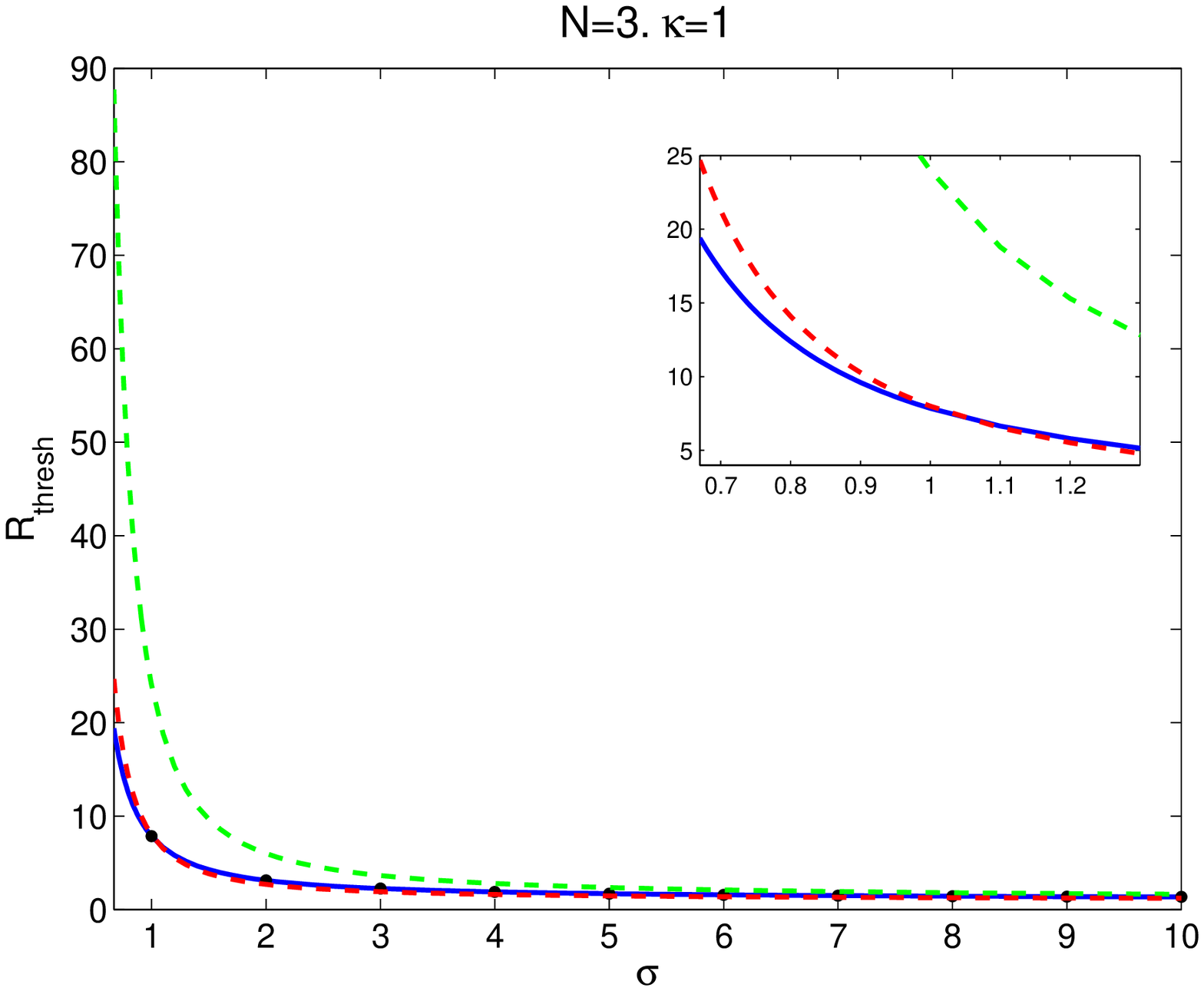}
    \end{tabular}
\caption{Numerical values for $\mathcal{R}_{\mathrm{thresh}}$ as a function of $\sigma\geq 2/N$ against its lower and upper estimation (\ref{res1})  for $\kappa_{\mathrm{crit}}=1$ (formula (\ref{res1t1})). (a) $N=1$, $\sigma\geq 2$, (b) $N=2$, $\sigma\geq 1$, (c) $N=3$, $\sigma\geq 2/3$. In all cases $\epsilon=1$. Green dashed line corresponds to the upper estimate, blue full line to the numerical $\mathcal{R}_{\mathrm{thresh}}$ and red dashed line to the lower estimate
The inset in (c) magnifies the discrepancy observed for the prediction of the lower estimate of (\ref{res1t1}) in the interval $\sigma\in (2/3,1)$. Black dots correspond to integer values of the nonlinearity exponent $\sigma$.} \label{fig1}
\end{center}
\end{figure}

Since the upper bound on (\ref{res1}) is exact, the estimates would have a full strength in applications, if the undetermined constant $\kappa_{\mathrm{crit}}$ could be easily determined, at least by a simple heuristic argument. For such a simple heuristic determination of the constant $\kappa_{\mathrm{crit}}$,  it looks natural to restrict to the case $\kappa\in\mathbb{Z}^+$, $\kappa\geq 1$. Then, the simplicity of the formula (\ref{res1}) suggests that the appropriate value of $\kappa$ can be determined by considering successive choices of $\kappa$. Setting
\begin{eqnarray*}
\mathcal{R}_{lb}=\left[\frac{\sqrt{2\kappa_{\mathrm{crit}}-1}}{\kappa_{\mathrm{crit}}}\cdot\frac{4N\epsilon(\sigma+1)}{2\sigma+1}
\right]^{\frac{1}{\sigma}},
\end{eqnarray*}
and rewriting (\ref{WGN2}) as
\begin{eqnarray*}
\mathcal{R}_{\mathrm{thresh}}=\left[\frac{(\sigma+1)\epsilon}{C_*}\right]^{\frac{1}{\sigma}},
\end{eqnarray*}
we observe that
\begin{eqnarray}
 \label{Rlims}
\lim_{\sigma\rightarrow\infty}\mathcal{R}_{lb}=\lim_{\sigma\rightarrow\infty}\mathcal{R}_{\mathrm{thresh}}=1.
\end{eqnarray}
{\em independently of the choice of $\kappa,\epsilon,N$. This behavior completely justifies that even the first choice $\kappa_{\mathrm{crit}}=1$ is valid for ``sufficiently large'' $\sigma$.}
The fist numerical study whose results are demonstrated in Figure \ref{fig1}, {\em examines the range of $\sigma>0$ on which this simplest choice $\kappa_{\mathrm{crit}}=1$ is valid, i.e. the validity of the formula}
\begin{eqnarray}
\label{res1t1}
\left[\frac{4N\epsilon(\sigma+1)}{2\sigma+1}
\right]^{\frac{1}{\sigma}}<R_{\mathrm{thresh}}<\left[4\epsilon N(\sigma+1)\right]^{\frac{1}{\sigma}}.
\end{eqnarray}
The green dashed line represents the theoretical upper estimate $\mathcal{R}_{ub}:=\left[4\epsilon N(\sigma+1)\right]^{\frac{1}{\sigma}}$, the  blue full line corresponds to the numerical $R_{\mathrm{thresh}}$ as a function of $\sigma\geq 2/N$ and the red dashed line represents the theoretical lower estimate $\mathcal{R}_{lb}$.
The first numerical study, not only reveals that the formula (\ref{res1t1})  is valid for the case $N=1,2$ but also of very good accuracy for $N=2$ and excellent for $N=3$ for $\sigma\geq 1$ with a discrepancy regarding the prediction of the lower bound $\mathcal{R}_{lb}$ appearing in the interval  $\sigma\in (2/3,1)$. \emph{In the light of the behavior (\ref{Rlims}), the choice $\kappa_{\mathrm{crit}}=1$ is satisfied for all $\sigma\geq 1$.}
\begin{figure}
\begin{center}
    \begin{tabular}{cc}
    \includegraphics[scale=0.45]{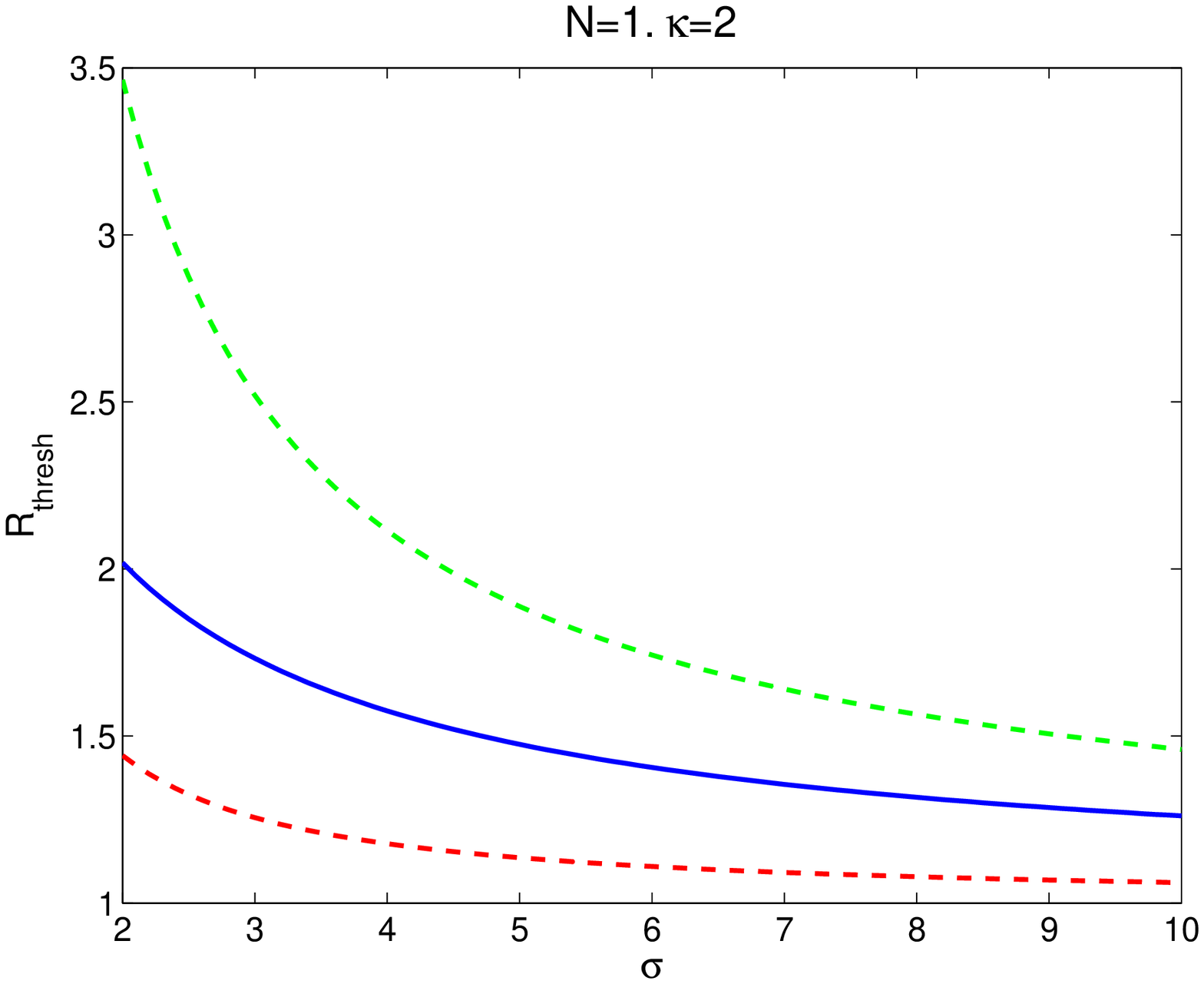} &
    \includegraphics[scale=0.45]{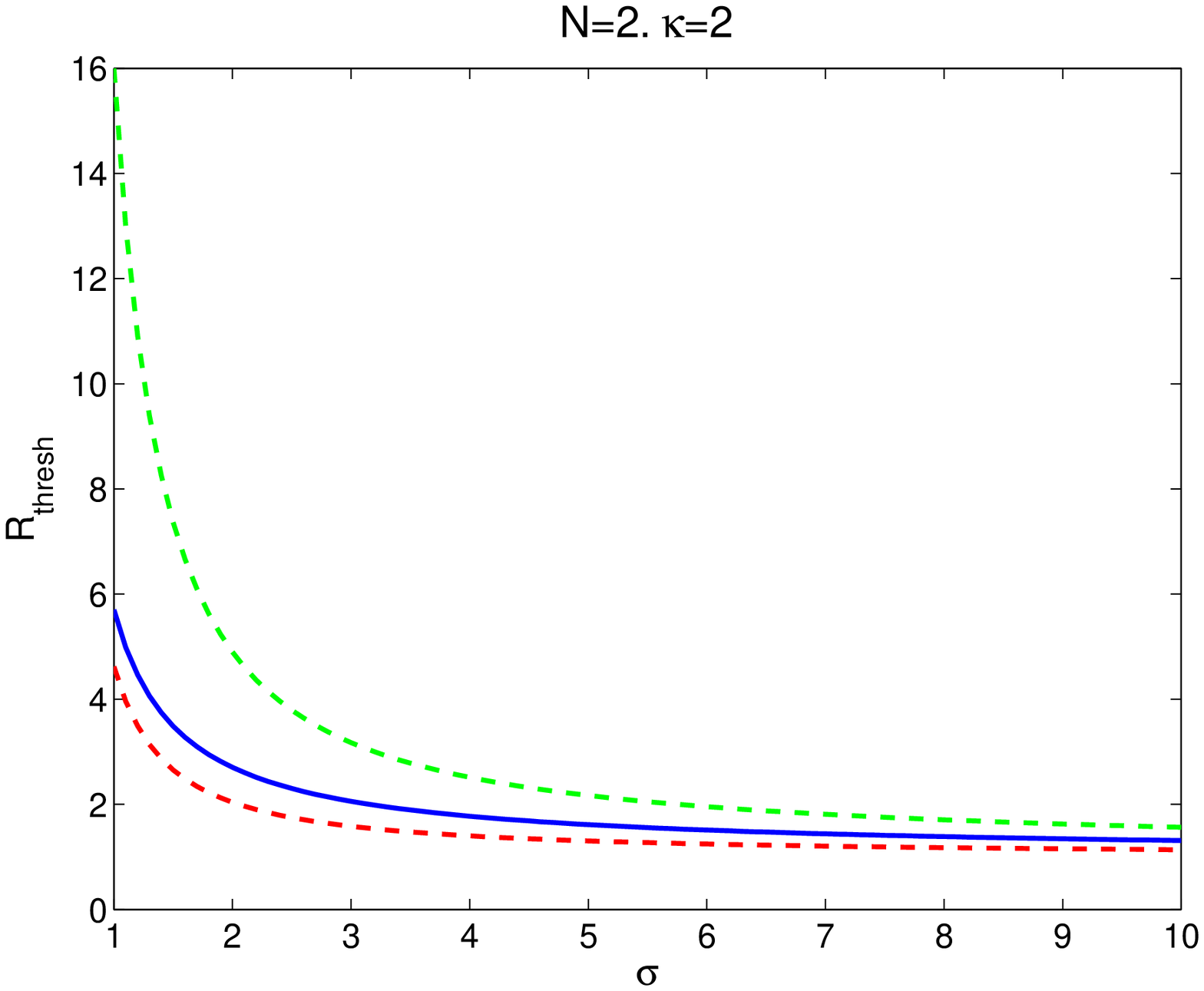} \\
    \includegraphics[scale=0.45]{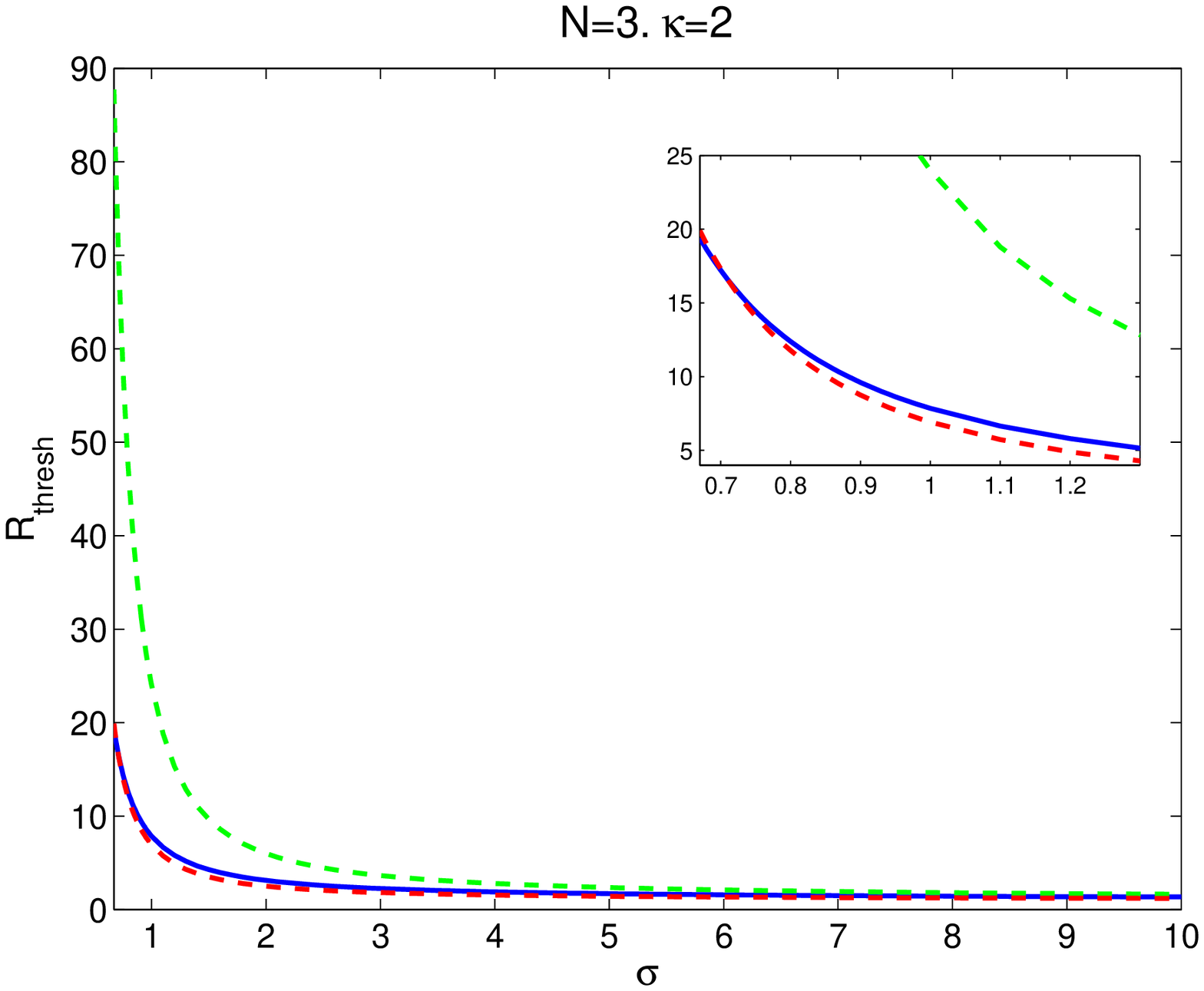}
    \end{tabular}
\caption{Numerical values for $\mathcal{R}_{\mathrm{thresh}}$ as a function of $\sigma\geq 2/N$ against its lower and upper estimation (\ref{res1})  for $\kappa_{\mathrm{crit}}=2$.
The inset in (c) magnifies the  discrepancy observed for the prediction of the lower estimate of (\ref{res2t2}) in the interval $\sigma\in (2/3,0.72)$ which is reduced in comparison with Figure \ref{fig1} (c). } \label{fig2}
\end{center}
\end{figure}
Motivated by the recent work of J. Dorignac, J. Zhou and D.K. Campbell \cite{DorZhouCam08}
which considers integer values of $\sigma\geq 2/N$ (represented by the black dots in the figures) it seems fair to state that the prediction of (\ref{res1t1}) is of particular usefulness for such nonlinearity exponents and lattice dimensions which are of main physical interest.
\begin{figure}
\begin{center}
    \begin{tabular}{cc}
    \includegraphics[scale=0.45]{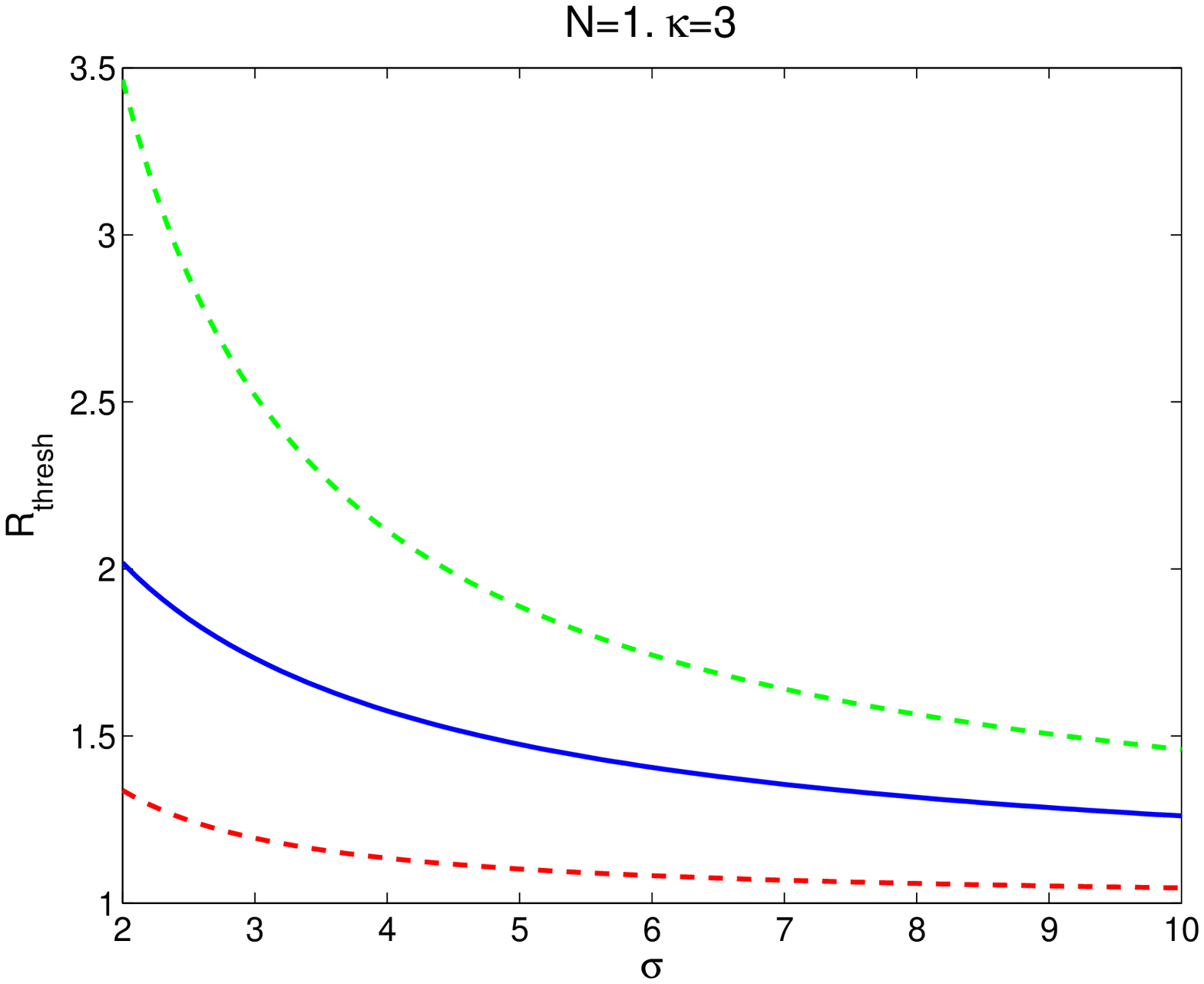} &
    \includegraphics[scale=0.45]{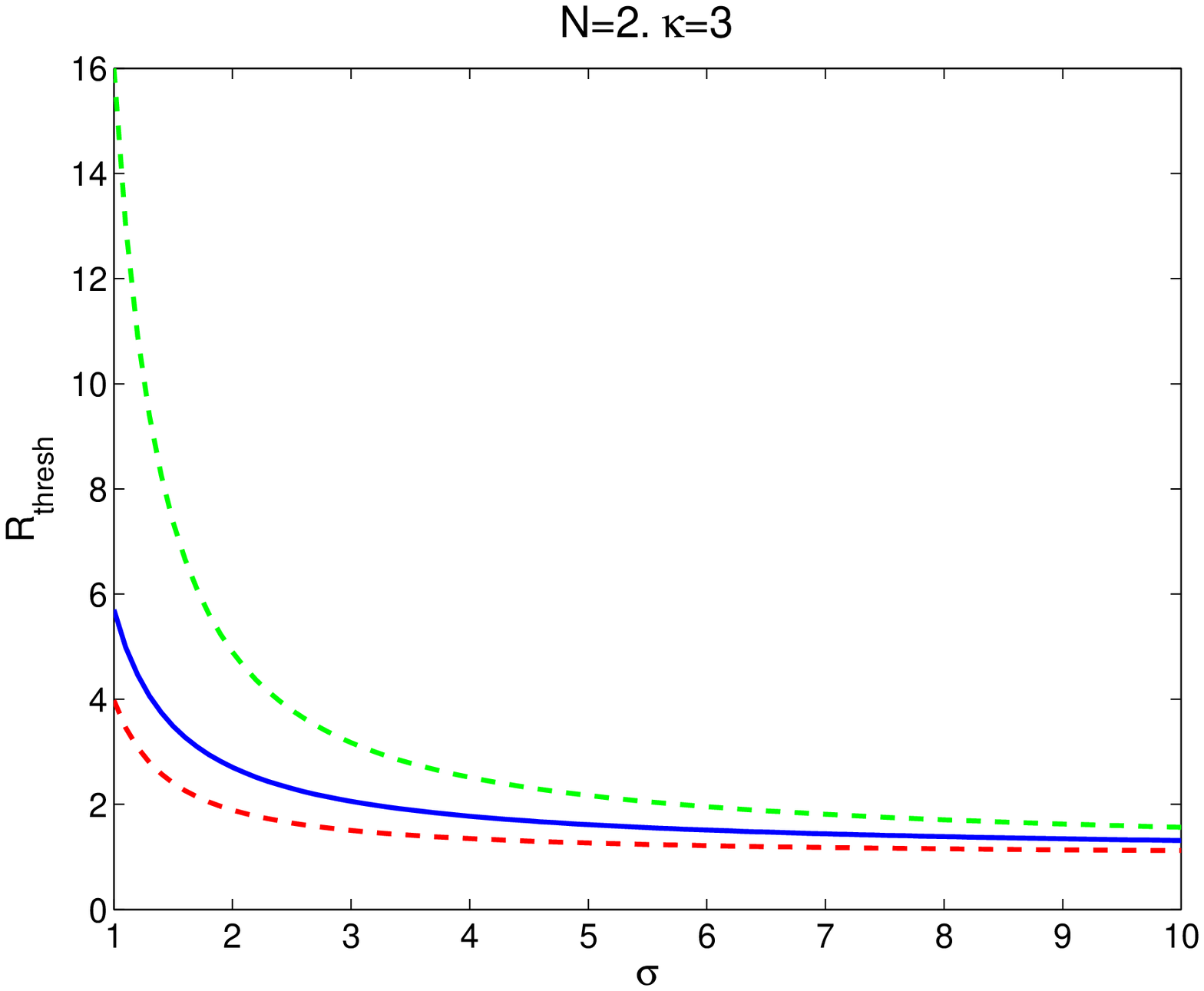} \\
    \includegraphics[scale=0.45]{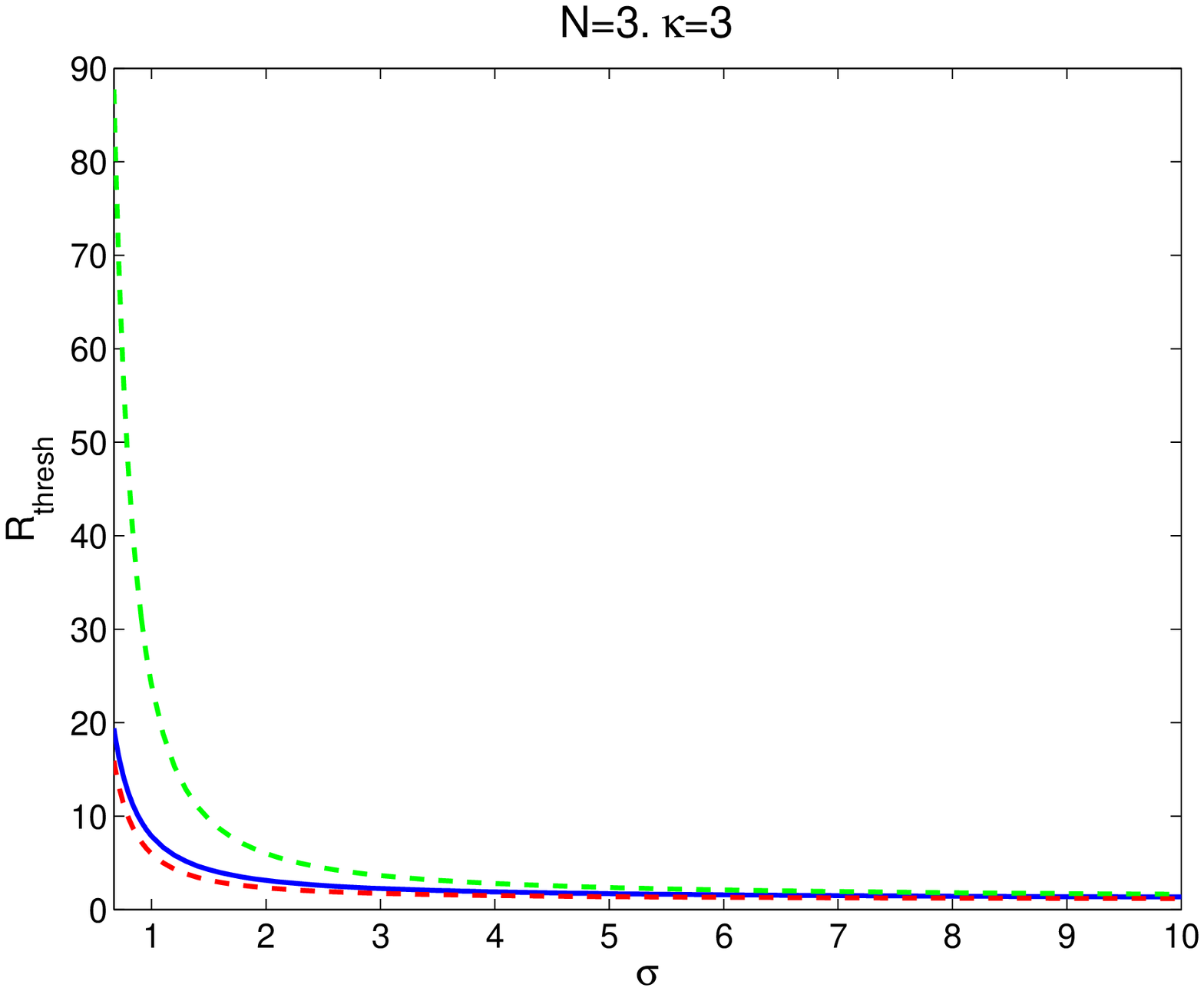}
    \end{tabular}
\caption{Numerical values for $\mathcal{R}_{\mathrm{thresh}}$ as a function of $\sigma\geq 2/N$ against its lower and upper estimation (\ref{res1})  for $\kappa_{\mathrm{crit}}=3$ (formula (\ref{res2t2a})).
The choice of $\kappa_{\mathrm{crit}}=3$ removes the discrepancies of Figures 1 (c) and 2 (c), suggesting the generalized formula (\ref{res3t3}) for the estimation of $\mathcal{R}_{\mathrm{thresh}}$.} \label{fig3}
\end{center}
\end{figure}
Seeking for the value of $\kappa_{\mathrm{crit}}$ which would remove the small discrepancy of (\ref{res1t1}) for $N=3$ and real values of $\sigma\geq 2/N$, our numerical findings in Figure \ref{fig2} verified that in the choice $\kappa_{\mathrm{crit}}=2$ this discrepancy is reduced to the interval $\sigma\in(2/3, 0.72)$ and it is completely removed for the choice of $\kappa_{\mathrm{crit}}=3$, as it is shown in Figure \ref{fig3}.  A summary of our findings for the cases $N=1,2,3$, suggests to restate Proposition \ref{prop1} taking into account the dependence of $\kappa_{\mathrm{crit}}$ on the dimension of the lattice: Letting $\kappa\in\mathbb{Z}^+$ and \emph{$N$ being fixed}, we observe that since $\lim_{\kappa\rightarrow\infty}\mathcal{R}_{\mathrm{min},2}(\kappa)=0$,we can always find $\kappa_{\mathrm{crit}}(N)\geq N$ such that
\begin{eqnarray}
 \label{res2t2}
\left[\frac{\sqrt{2\kappa_{\mathrm{crit}}(N)-1}}{\kappa_{\mathrm{crit}}(N)}\cdot\frac{4N\epsilon(\sigma+1)}{2\sigma+1}
\right]^{\frac{1}{\sigma}}<R_{\mathrm{thresh}}<\left[4\epsilon N(\sigma+1)\right]^{\frac{1}{\sigma}},\,\mbox{for all}\;\; 1\leq N\leq \kappa_{\mathrm{crit}}(N).
\end{eqnarray}
With the rigorously valid estimates (\ref{res2t2}) at hand, the numerical study  for the cases $N=1,2,3$ suggest that when $N=3$ it is justified to consider $\kappa_{\mathrm{crit}}(N)=3$ and that
\begin{eqnarray}
 \label{res2t2a}
\left[\frac{\sqrt{5}}{3}\cdot\frac{4N\epsilon(\sigma+1)}{2\sigma+1}
\right]^{\frac{1}{\sigma}}<R_{\mathrm{thresh}}<\left[4\epsilon N(\sigma+1)\right]^{\frac{1}{\sigma}},\;\;\mbox{for all}\;\; 1\leq N\leq 3.
\end{eqnarray}

Actually the numerical study and especially the collection of Figures \ref{fig1} (a), \ref{fig2} (b) and \ref{fig3} (c) justify the validity of the formula
\begin{eqnarray}
 \label{res3t3}
\left[\frac{\sqrt{N-1}}{N}\cdot\frac{4N\epsilon(\sigma+1)}{2\sigma+1}
\right]^{\frac{1}{\sigma}}<R_{\mathrm{thresh}}<\left[4\epsilon N(\sigma+1)\right]^{\frac{1}{\sigma}},\;\;\mbox{for all}\;\; 1\leq N\leq 3,
\end{eqnarray}
which is of valuable accuracy for $N=2,3$.
The estimates (\ref{res2t2a}), (\ref{res3t3}) have the advantage of removing the small discrepancy of (\ref{res1t1}) observed in the case $N=3$, for real $\sigma\geq 2/N$. However we believe that all the above  formulas derived by a simple heuristic implementation of Proposition \ref{prop1}, serve as a very satisfactory analytical estimation of the excitation threshold in the cases of $\sigma, N$ which are of physical significance.

\section{Conclusions}
In this work, we have determined analytical upper and  lower estimates on the excitation threshold for breathers in  $N$--dimensional DNLS lattices. Numerical calculations show that, in cases studied, the theoretical bound is close to the  true threshold providing useful analytical expressions to determine analytical energy activation of breathers in these systems. On the other hand, extensions of previous results to more general situations, as DNLS systems with impurities, are currently under investigation and will be reported in future publications.


\ack

FP and JC acknowledge financial support from the MECD project FIS2008-04848.

\section*{References}

\bibliographystyle{amsplain}

\end{document}